\documentclass[10pt,journal,compsoc]{IEEEtran}
\ifCLASSOPTIONcompsoc
  \usepackage[nocompress]{cite}
\else
  \usepackage{cite}
\fi

\usepackage{url}
\usepackage{graphicx}
\usepackage{amssymb}

\usepackage{dblfloatfix}

\usepackage{pgfplots}
\usepgfplotslibrary{colorbrewer}
\pgfplotsset{width=7.75cm, compat = 1.8,
             cycle list/Set2} 
\usetikzlibrary{pgfplots.statistics, pgfplots.colorbrewer} 
\usepackage{pgfplotstable}

\usepackage{dsfont}
\def\RSet{\mathds{R}}

\def\transpose#1{#1^\intercal}

\newcommand*{\E}{\mathbb{E}}

\usepackage[draft,inline,nomargin]{fixme}

\fxsetup{theme=colorsig}
\definecolor{fxnote}{rgb}{0.8000,0.5000,0.0000}
\definecolor{fxerror}{rgb}{0.9900,0.0000,0.0000}

\usepackage{siunitx}
\begin{document}

\title{Off the Beaten Track:
Using Deep Learning to Interpolate Between Music Genres}

\author{Tijn~Borghuis, Alessandro~Tibo, Simone~Conforti,
  Luca~Canciello, Lorenzo~Brusci, Paolo~Frasconi
\IEEEcompsocitemizethanks{\IEEEcompsocthanksitem T. Borghuis is with Eindhoven University of Technology (NL).\protect\\
E-mail: v.a.j.borghuis@tue.nl
\IEEEcompsocthanksitem A. Tibo is with University of Florence (IT).\protect\\
E-mail: alessandro.tibo@unifi.it
\IEEEcompsocthanksitem S. Conforti is with University of Basel (CH) and Conservatorio G.F. Ghedini di Cuneo (IT).\protect\\
E-mail: simone.conforti@unibas.ch
\IEEEcompsocthanksitem L. Canciello is a freelance composer and sound designer and was with Musica Combinatoria, Krakow (PL) during the development of this project.\protect\\
E-mail: info@lucacanciello.com
\IEEEcompsocthanksitem L. Brusci is with Musica Combinatoria, Krakow (PL).\protect\\
E-mail: lorenzo.brusci@musi-co.com
\IEEEcompsocthanksitem P. Frasconi is with University of Florence (IT).\protect\\
E-mail: see http://ai.dinfo.unifi.it/paolo
}
}

\date{\today}

\maketitle

\begin{abstract}
  We describe a system based on deep learning that generates drum
  patterns in the electronic dance music domain. Experimental results
  reveal that generated patterns can be employed to produce musically
  sound and creative transitions between different genres, and that
  the process of generation is of interest to practitioners in the
  field.
\end{abstract}

Automatic music generation is a fast growing area with applications in
diverse domains such as gaming~\cite{plans_experience-driven_2012},
virtual environments~\cite{casella_magenta:_2001}, and entertainment
industry~\cite{nakamura_automatic_1994}. A thorough account of goals
and techniques is presented in~\cite{pasquier_introduction_2017}.
Electronic dance music (EDM) is one of the domains where automatic
generation appears to be particularly promising due to its heavily
constrained repetitive structure, characterized by clearly defined
stylistic
forms~\cite{eigenfeldt_evolving_2013}.  In this domain, a
typical task of ``traditional'' Disk-Jockeys (DJs) operating in dance
clubs and radio stations is to obtain seamless transitions between the
consecutive tracks of a given playlist. When the individual song
tracks are entirely pre-recorded, the main tools available to DJs are
a combination of an accurate synchronization of beats-per-minute (BPM)
via beatmatching, and a perceptually smooth crossfading (i.e.\,
gradually lowering the volume of one track while increasing the volume
of the other track), occasionally with the help of equalizers and
other effects such as reverbs, phasers, or delays, which are commonly
available in commercial mixing consoles. However, since the early
1990's, it has become increasingly common for DJs to take advantage of
samplers and synthesizers that can be used to generate (compose)
on-the-fly novel musical parts to be combined with the existing
pre-recorded material. Artists such as DJ Shadow and DJ Spooky have
indeed demonstrated that the line demarcating DJing and electronic
music composition can be extremely blurry~\cite{katz_capturing_2004}.

In this paper, we explore the automation of track transitioning when
the consecutive tracks belong to different genres.
Previous attempts to automate track transitioning are
limited to time-stretching and crossfading~\cite{cliff_hang_2000},
essentially mimicking the human work of a traditional DJ.\ While these
approaches can effectively achieve the goal of a seamless mix between
pre-recorded tracks (particularly when the tracks stay within the same
musical genre), they hardly fit the contemporary scenario where DJs
%
%
and composers are seeking artistically interesting results also by
using in a creative way transitions spanning different genres.
We thus advocate a
radically different perspective where novel musical material is
automatically generated (composed) by the computer in order to smoothly
transition from one genre to another.  The hope is that by exploring
this smooth musical space in an automatic fashion, we can create
materials that are useful to musicians as novel elements in
composition and performance.
In this new approach, we thus take transitions be compositional devices 
in their own right. Depending on length, transitions serve
different functions in dance music: they could be momentary sound
effects on a short time scale, or comparable to frequency sweeps on
longer time scales, or on even longer scales to act as an
automatically generated foundation on top of which musical layers can
be added, both in live performance and post-production. 

Building a general interpolation tool that encompasses the whole set
of instruments is beyond the scope of this paper.  Genre in the domain
of EDM is mainly determined by the basic rhythm
structure of drum instruments. Hence, by taking these drum patterns as
the musical material for an experiment, the complexity associated with
other aspects such as harmony, melody and
timbre~\cite{boulanger-lewandowski_modeling_2012},
that are relevant for genre in other musical domains, can be
sidestepped.  At the same time, the domain remains sufficiently
complex and suitable for significant real-world usage by music
professionals such as DJs, producers, and electronic musicians.

To investigate the potential of deep learning for transitioning
between dance music genres, we designed a learning system based on
variational autoencoders~\cite{kingma_auto-encoding_2013}, we trained
it on a dataset of rhythm patters from different genres that we
created, and then we asked it to produce \textit{interpolations}
between two given rhythm patterns. Such transitions consist of a
sequence of rhythm patterns, starting from a given pattern from one
genre and ending in a given goal pattern (possibly from another
genre). The connecting patterns in between are new rhythm patterns
generated by the trained system itself. Along the same vein, we also
developed an autonomous drummer that smoothly explores the EDM drum
pattern space by moving in the noise space of a generative adversarial
network~\cite{goodfellow_generative_2014}.  We constructed an
experimental software instrument that allows practitioners to create
interpolations, by embedding the learning system in Ableton Live, a
music software tool commonly used within the EDM production community.
Finally, the proposed process for creating interpolations and the
resulting musical materials were thoroughly evaluated by a set of
musicians especially recruited for this research.

\section*{Generative Machine Learning}
Many machine learning applications are concerned with pattern
recognition problems in the supervised setting.  Recently, however, a
rather different set of problems has received significant
attention, where the goal is to \textit{generate} patterns rather than
\textit{recognize} them.  Application domains are numerous and diverse
and very often involve the generation of data for multimedia
environments. Examples include natural
images~\cite{radford_unsupervised_2015},
videos~\cite{vondrick_generating_2016},
paintings~\cite{elgammal_can:_2017},
text~\cite{yu_seqgan:_2016},
and music~\cite{yang_midinet:_2017,yu_seqgan:_2016,boulanger-lewandowski_modeling_2012}.

Pattern generation is closely related to unsupervised learning, where
a dataset $\{x^{(1)},\dots,x^{(n)}\}$ of patterns, sampled from an
unknown distribution $p$, is given as input to a learning algorithm
whose task is to estimate $p$ or to extract useful information about
the structure of $p$ such as clusters (i.e.\, groups of similar
patterns) or support (i.e.\, regions of high density, especially when
it consists of a low-dimensional manifold). In pattern generation,
however, we are specifically interested in \textit{sampling} new
patterns from a distribution that matches $p$ as well as possible. Two
important techniques for pattern generation are generative adversarial
networks and (variational) autoencoders, briefly reviewed in the
following.

\subsection*{Generative Adversarial Networks}
Generative Adversarial Networks
(GANs)~\cite{goodfellow_generative_2014} consist of a pair of neural
networks: a \textit{generator}, $\mathcal{G}:\RSet^d\mapsto\RSet^m$,
parameterized by weights $w_g$, and a \textit{discriminator},
$\mathcal{D}:\RSet^m\mapsto \{0,1\}$, parameterized by weights
$w_d$. The generator receives as input a vector $z\in\RSet^d$ sampled
from a given distribution $q$ and outputs a corresponding pattern
$\mathcal{G}(z)\in\RSet^m$. We can interpret $z$ as a low-dimensional
code for the generated pattern, or as a tuple of coordinates within
the manifold of patterns.  The discriminator is a binary classifier,
trained to separate \textit{true} patterns belonging to the training
dataset (positive examples) from \textit{fake} patterns produced by
the generator (negative examples). Training a GAN is based on an
adversarial game where the generator tries to produce fake patterns
that are as hard to distinguish from true patterns as possible, while
the discriminator tries to detect fake patterns with the highest
possible accuracy. At the end of training we hope to reach a game
equilibrium where the generator produces realistic patterns as
desired. The discriminator is no longer useful after
training. Equilibrium is sought by minimizing the following objective
functions, for the discriminator and for the generator, respectively:
\begin{IEEEeqnarray}{lCl}
  \label{eq:discriminator-obj}
  J_d(w_d) &=& \E_{x\sim p} \left[ L(\mathcal{D}(x),1) \right] +
  \E_{z\sim q} \left[ L(\mathcal{D}(\mathcal{G}(z)),0) \right]\\
  \label{eq:generator-obj}
  J_g(w_g) &=& \E_{z\sim q} \left[ L(\mathcal{D}(\mathcal{G}(z)),1) \right]
\end{IEEEeqnarray}
where $L$ denotes the binary cross-entropy loss, $q$ can be either a
uniform distribution on a compact subset of $\RSet^d$ or,
alternatively, a Gaussian distribution with zero mean and unit
variance. Since $p$ is not
accessible, the expectation in Eq.~(\ref{eq:discriminator-obj}) is
replaced by its empirical value on the training sample. In practice,
fake data points are also sampled.  Optimization typically proceeds by
stochastic gradient descent or related algorithms where a balanced
minibatch of real and fake examples is generated at each optimization
step.

\subsection*{Autoencoders and Variational Autoencoders}
Autoencoders also consist of a pair of networks: an encoder,
$\mathcal{E}$, parameterized by weights $w_e$, that maps an input
pattern $x\in\RSet^m$ into a latent code vector
$z=\mathcal{E}(x)\in\RSet^d$, and a decoder, $\mathcal{D}$,
parameterized by weights $w_d$, mapping latent vectors $z\in\RSet^d$
back to the pattern space $\RSet^m$. In this case, the two networks
are stacked one on the top of the other to create a composite function
$\mathcal{D} \circ \mathcal{E} :\RSet^m\mapsto\RSet^m$, and the
overall model is trained to reproduce its own inputs at the
output. Since typically $d\ll m$, the model is forced to develop a
low-dimensional representation that captures the manifold of the
pattern associated with the data distribution $p$. Training is
performed by minimizing the objective
\begin{IEEEeqnarray}{lCl}
  \label{eq:autoencoder-obj}
  J(w_e,w_d) &=& \E_{x\sim p} \left[ L(\mathcal{D}(\mathcal{E}(x)),x) \right]
\end{IEEEeqnarray}
where the parameters $w_e$ and $w_d$ are optimized jointly and $L$ an
appropriate reconstruction loss.

Variational autoencoders (VAEs)~\cite{kingma_auto-encoding_2013} also
consist of an encoder and a decoder, but they bear a probabilistic
interpretation. To generate a pattern, we first sample a vector
$z\in\RSet^d$ from a prior distribution $p(z)$ (usually a multivariate
Gaussian with zero mean and unit variance), and we then apply $z$ as
input to the decoder, in order to obtain $p(x|z)$. The encoder in this
case produces an approximation $q(z|x)$ to the intractable posterior
$p(z|x)$. Specifically, $q(z|x)$ is a multivariate Gaussian
whose mean $\mu(x)$ and diagonal covariance $\sigma(x)$ are computed
by the encoder network $\mathcal{E}$ receiving a pattern $x$ as
input. A VAE is then trained to minimize the difference between the
Kullback-Leibler divergence
\begin{IEEEeqnarray}{lCl}
  \label{eq:vae-loss1}
  \mathrm{KL}(q(z)||p(z)) & = & \int q(z) \log \frac{p(z)}{q(z)} dz\\
  & = & \frac{1}{2} \sum_{j=1}^d
  \left( 1 + \log \sigma^2_j(x) - \mu^2_j(x) - \sigma^2_j(x) \right) \nonumber
\end{IEEEeqnarray}
and the log conditional likelihood
\begin{IEEEeqnarray}{lCl}
  \label{eq:vae-loss2}
  \log p(x|z) = - \E_{x\sim p} \left[ L(x,\mathcal{D}(z)) \right].
\end{IEEEeqnarray}

\subsection*{Deep and Recurrent Networks}
\begin{figure}[!t]
  \begin{center}
    \includegraphics[width=.49\textwidth]{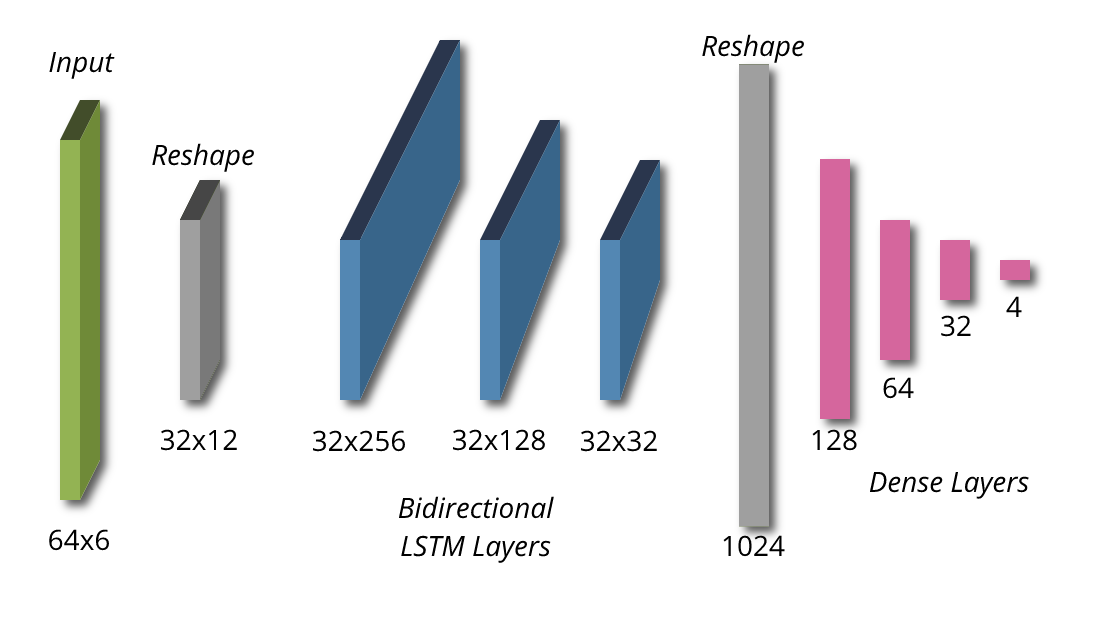}
    ~\\
    ~\\
    \includegraphics[width=.3\textwidth]{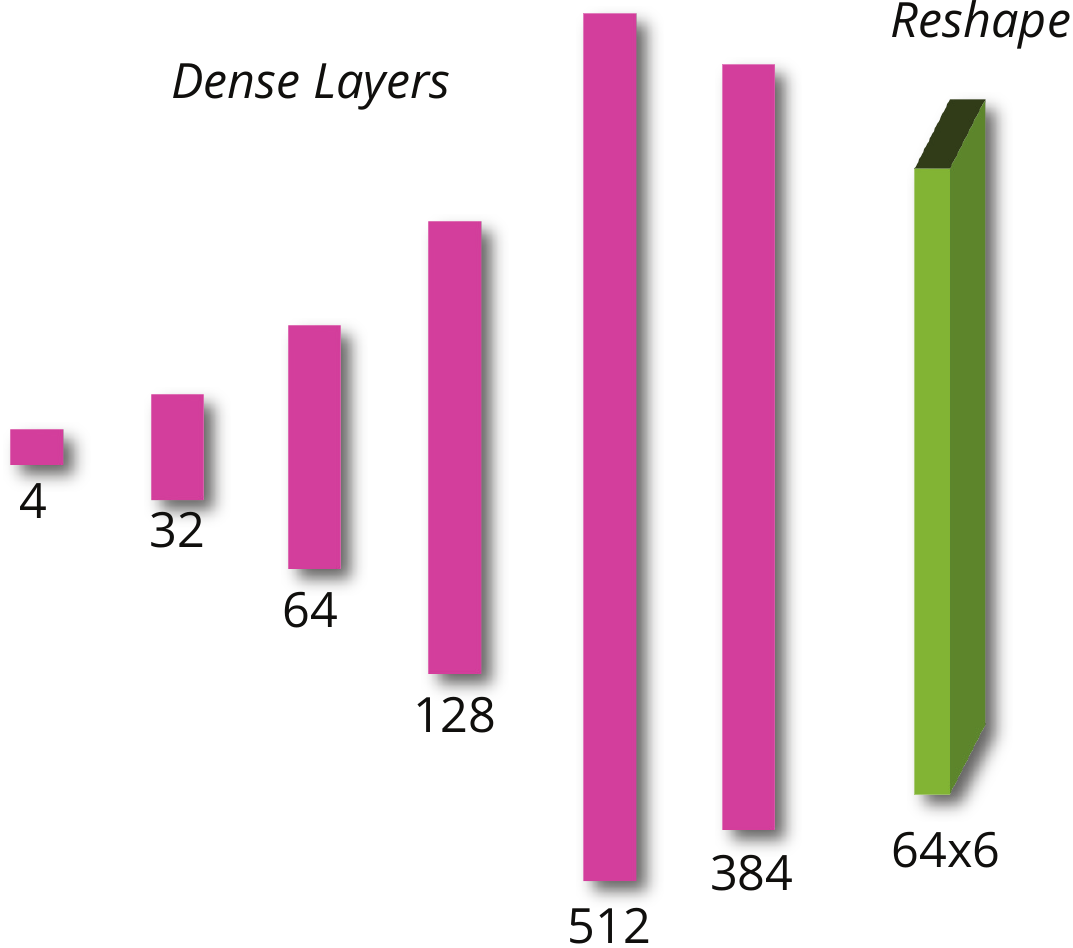}
    \caption{\label{fig:architecture}
      Architecture of the variational autoencoder used to
      interpolate drum patterns. Top: Encoder; Bottom: Decoder.}
  \end{center}
\end{figure}

All the above networks (generator and discriminator for GANs, decoder
and encoder for VAEs) can be constructed by stacking several
neural network layers. In particular, our encoder for the VAE was
based on three bidirectional long-short-term-memory
(LSTM)~\cite{hochreiter_long_1997} recurrent layers with tanh
nonlinearities, followed by four fully connected layers with ReLU
nonlinearities, ending in a representation of size $d=4$. LSTM
layers were used to capture the temporal structure of the data and, in
particular, the correlations among note-on MIDI events within a drum
pattern. Convolutional layers could have also been employed and we
found that they produce similar reconstruction errors during
training. We developed a slight aesthetic preference towards LSTM
layers in our preliminary listening sessions during the development of
the VAE, although differences compared to convolutional layers were
not very strong.
The decoder simply consisted of five fully connected
layers with ReLUs. We used logistic units on the last layer of the
decoder and a binary cross-entropy loss for comparing reconstructions
against true patterns, where MIDI velocities were converted into
probabilities by normalizing them in [0,1].  Details on the
architecture are visible in Figure~\ref{fig:architecture}.

The discriminator and the generator networks for the GAN had
essentially the same architectures as the encoder and the decoder for
the VAE, respectively, except of course the GAN discriminator terminates
with a single logistic unit and for the VAE we used a slightly
smaller (two-dimensional) noise space, in order to exploit the
``swirling'' explorer described below in the ``autonomous
drumming'' subsection.

\subsection*{Electronic Dance Music Dataset}

One of the authors, who is a professional musician, used his in-depth
knowledge of EDM to compose a collection of drum patterns
representative of three genres: Electro, Techno, and Intelligent Dance
Music (IDM).  In all patterns, the following six instruments of a
Roland TR-808 Rhythm composer drum machine were used: bass drum, snare
drum, closed hi-hat, open hi-hat, rimshot, and cowbell. The TR-808
(together with its sisters TR-606 and TR-909), was integral to the
development of electronic dance music and these six instrument sounds
are still widely used in EDM genres today which makes them suitable
for our interpolation approach. All patterns are one measure (4 bars)
long, and quantized to 1/16th note on the temporal scale. At the
intended tempo of 129 BPM, it takes 7.44s to play one
measure. Patterns were constructed with the help of the Ableton Live
music production software, and delivered in the form of standard
MIDI files. After checking for duplicates, a data set consisting of
1782 patterns resulted, which is summarized in
Table~\ref{tab:patterns}.

Each drum pattern was represented as a two-dimensional array whose
first and second axes are associated with the six selected drum
instruments and the temporal position at which a MIDI note-on event
occurs, respectively. Note durations were not included in the
representation as they are irrelevant for our choice of percussive
instruments.  The duration of four measures results in a $6\times 64$
array for each pattern. Values (originally in the integer range [0,127], then
normalized in [0,1]) correspond to MIDI velocities and were used
during dataset construction mainly to represent dynamic accents or
ghost (echoing) notes that may be present in some musical styles. In
our representation, a zero entry in the array indicates the absence of
a note-on event.

\begin{figure*}[!t]
  \begin{center}
    \includegraphics[width=.99\textwidth]{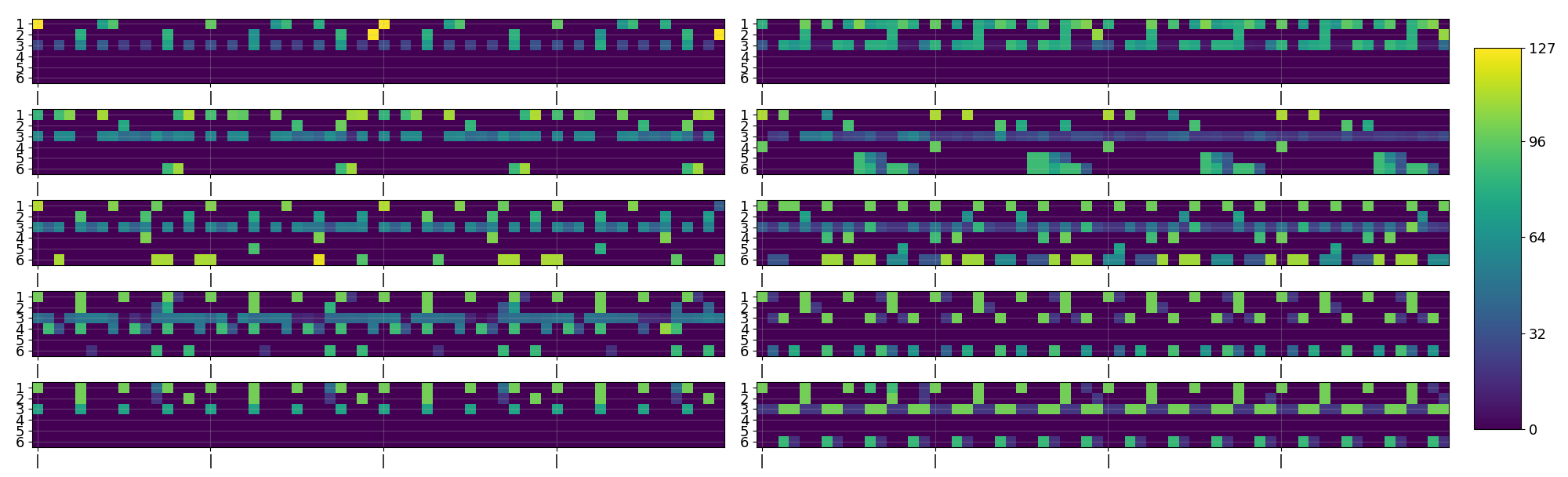}
    \caption{\label{fig:patterns} Ten sample drum patterns in
      the EDM dataset. Instruments from the top are
      (1): bass drum, (2): snare drum, (3): closed hi-hat, (4): open
      hi-hat, (5): rimshot, (6): cowbell. Pixel intensities correspond
      to MIDI velocities. Top row: Electro-Funk; mid two rows: IDM;
      bottom two rows: Techno.}
  \end{center}
\end{figure*}

\begin{table}[htp]
  \caption{Electronic Dance Music Dataset}
  \label{tab:patterns}
  \begin{center}
    \begin{tabular}{lS[table-format = 5.0]S[table-format = 5.0]l}
      {\bf Style} & {\bf \# of patterns} & \multicolumn{2}{c}{\bf Playing time} \\
      \hline
      IDM     &  608 &  4,525s &(1h 15m 25s) \\
      Electro &  690 &  5,135s &(1h 25m 35s) \\
      Techno  &  484 &  3,602s &(1h 0m 2s) \\
      \hline
      Total   & 1,782 & 13,261s &(3h 41m 1s) \\
    \end{tabular}
  \end{center}
\end{table}

\section*{Generating interpolations}

\begin{figure}[!t]
  \begin{center}
    \includegraphics[width=.5\textwidth]{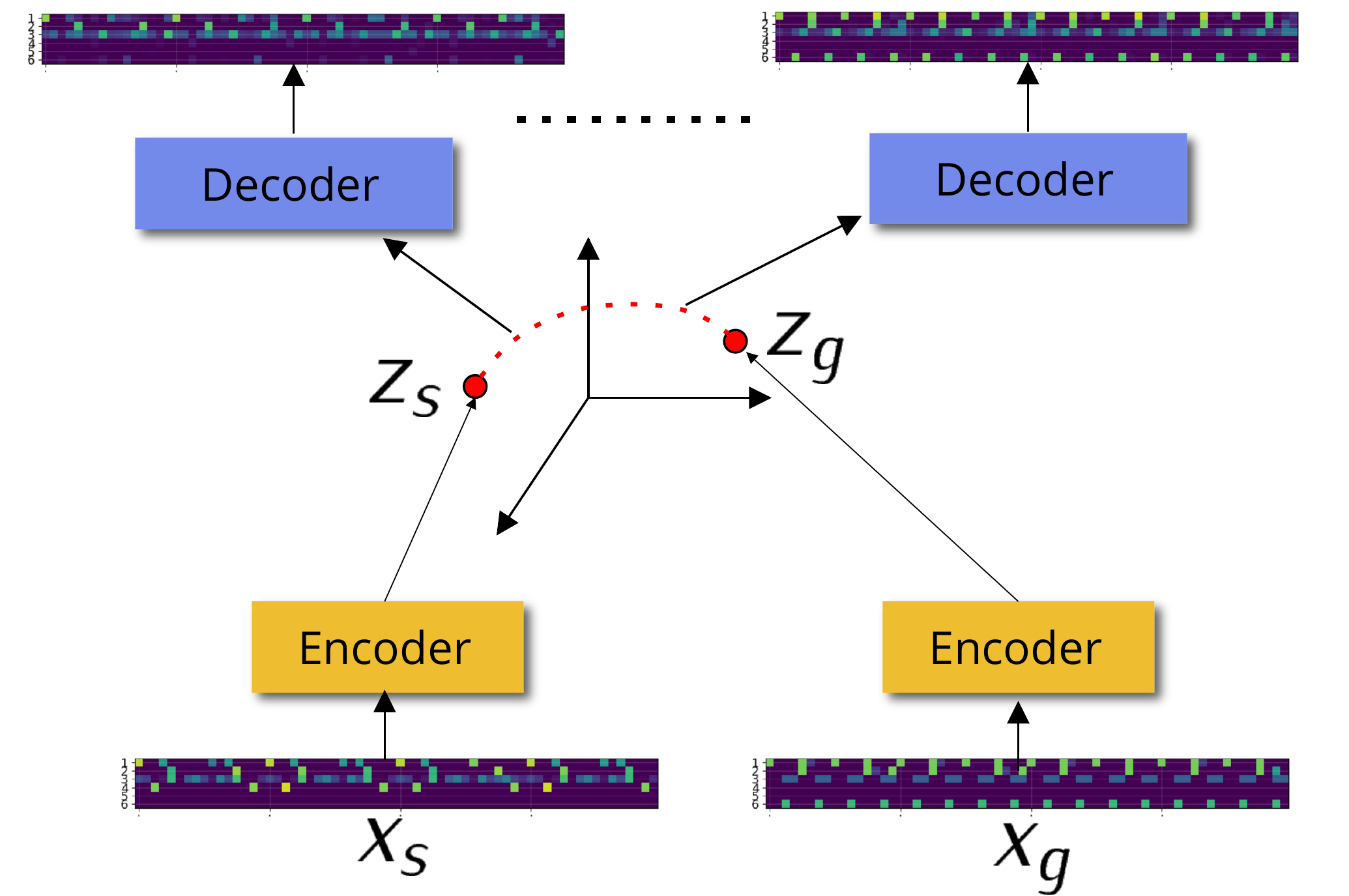}
    \caption{\label{fig:interpolations}
      Building transitions by
      interpolating drum patterns in their representation space.}
  \end{center}
\end{figure}

Both techniques discussed above were used to generate sequences of
drum patterns that interpolate between genres.

\subsection*{Using VAEs for start-goal interpolations}
When using VAEs, it is straightforward to create an interpolation
between a starting pattern $x_s$ and a goal pattern $x_g$ as follows
(see also Figure~\ref{fig:interpolations}):
\begin{enumerate}
\item Apply the encoder $\mathcal{E}$ to the endpoint patterns to obtain the
  associated coordinates in the manifold space of the autoencoder:
  $z_s=\mathcal{E}(x_s)$ and $z_g=\mathcal{E}(x_g)$;
\item For a given interpolation length, $L$, construct a sequence of
  codes in the manifold space:
  $\langle z_0=z_s, z_1, \dots, z_L=z_g\rangle$
\item Apply the decoder $\mathcal{D}$ to each element of this
  sequence, to obtain a sequence of patterns:
  $\langle \mathcal{D}(z_0), \dots, \mathcal{D}(z_L)\rangle$; note
  that (unless the autoencoder underfits the dataset)
  $\mathcal{D}(z_0)\approx x_s$ and $\mathcal{D}(z_L)\approx x_g$.
\end{enumerate}

\subsection*{Linear and spherical interpolation}
In the case of linear interpolation (LERP), the sequence of codes is
defined as
\begin{equation}
\label{eq:lerp}
z_i = (1-\mu_i) z_s + \mu_i z_g
\end{equation}
for $\mu_i=i/L$, $i=0, \dots L$.  In the case of spherical
interpolation (SLERP), the sequence is
\begin{equation}
  \label{eq:slerp}
  z_i = \frac{z_s \sin(\theta(1.0-\mu_i)) + z_g \sin(\theta\mu_i)}{\sin(\theta)}
\end{equation}
where $\theta=\arccos(\frac{\transpose{h}_s z_g}{\|z_s\|\|z_g\|})$.
\cite{white_sampling_2016} offers a thorough discussion of the
benefits of SLERP in the case of image generation. We found that SLERP
interpolations produced musically more adventurous and expressive
results and thus we used them in our experimental evaluation.

\subsection*{Crossfading vs.\ interpolation in the representation space}

\begin{figure*}[!t]
  \begin{center}
    \includegraphics[width=0.63\textwidth]{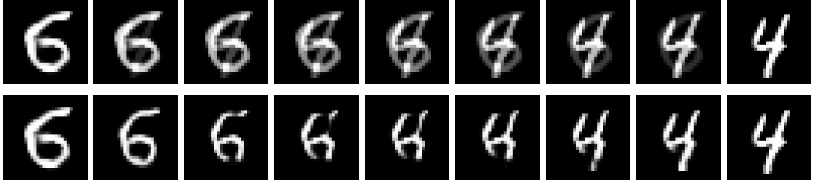}
    \caption{\label{fig:interpolating-vs-xfading} Top: Interpolation
      in the pattern space (i.e., crossfading) between two MNIST characters; Bottom:
      interpolation in the representation space.}
  \end{center}
\end{figure*}

We remark the significance of performing the interpolation in the
representation space: rather than generating a weighted average of two
patterns (as it would happen with crossfading, which consists of a
linear combination as in Eq.~\ref{eq:lerp} but using identity
functions instead of $\mathcal{E}$ and $\mathcal{D}$), we generate at
each step $i$ a \textit{novel drum pattern} from the learned
distribution. To help the reader with a visual analogy, we show in
Figure~\ref{fig:interpolating-vs-xfading} the difference between
interpolation in pattern space (crossfading) and in representation
space using two handwritten characters from the MNIST dataset.

\subsection*{Pattern novelty}
\begin{figure}
  \begin{center}
    \begin{tabular}{cc}
      \includegraphics[width=0.45\textwidth]{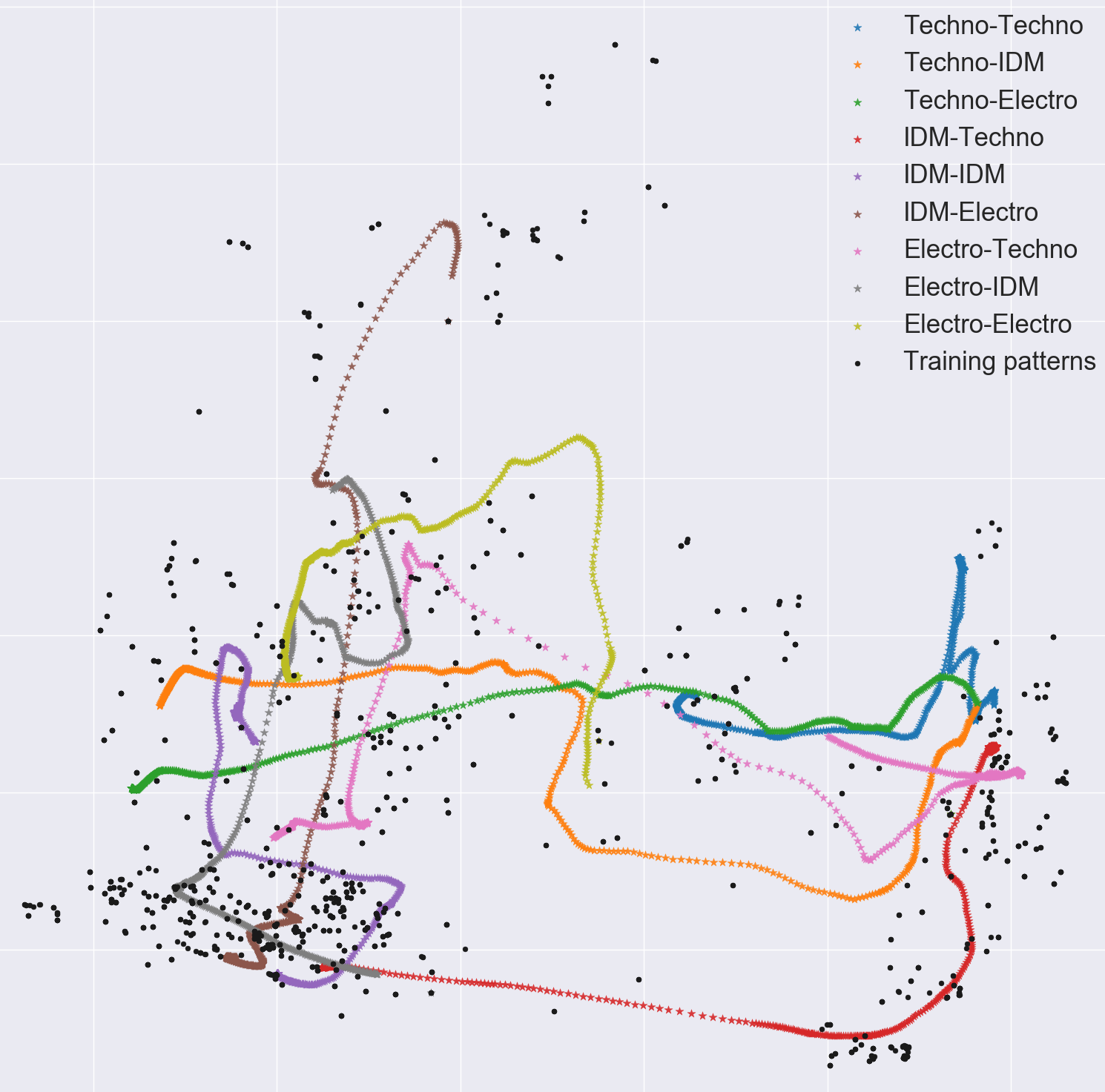}
    \end{tabular}
    \caption{PCA plot of training data (black dots) and a set of
      possible start-goal interpolations obtained with a deep LSTM VAE
      (labeled by the genres of the start and goal patterns).}
   \label{fig:PCA}
  \end{center}
\end{figure}

A quantitative measure of quality and novelty of patterns generated by
models such as VAEs or GANs is not readily available. We observed
however that several of the patterns produced by interpolating between
start and goal patterns in our dataset are genuinely new. In
Figure~\ref{fig:PCA} we visualize the result of two-dimensional principal components analysis
(PCA) showing all training set patterns and those generated by
interpolating between a subset of them. It can be seen that
trajectories tend to respect the distribution of the training data but
include new data points, showing that novel patterns are indeed
generated in the transitions.

\subsection*{A software instrument for start-goal interpolations}
\label{subsec:interface}
The trained VAE (in the form of a Tensorflow model) was embedded as a
plugin in Ableton Live Suite 9 for Mac OS, a program that is widely
used by performing and producing musicians in EDM, and that enables
the construction of software instruments via the programming
environment \textit{Max for Live}.
During performance, musicians first specify a start and a goal pattern
(chosen from the dataset), and the length of the interpolation. This
can be conveniently done within the Live user interface. The
controller (a small Python script) then produces the required sequence
of patterns using the VAE and the resulting MIDI notes are sent to
\textit{Live} to be rendered in audio with a user-specified
soundset. The whole process is fast enough for real-time usage.

\subsection*{Using GANs for autonomous drumming}
\begin{figure}
  \begin{center}
    \includegraphics[width=0.35\textwidth]{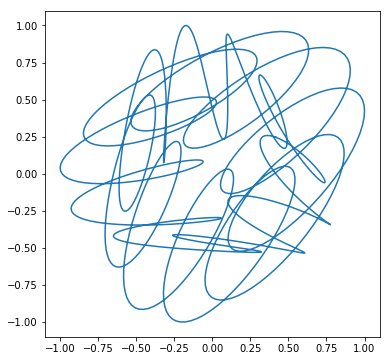}
    \caption{Swirl in GAN noise space associated with Eq.~\ref{eq:swirl}.}
   \label{fig:swirl}
  \end{center}
\end{figure}

In the case of GANs, Step 1 of the procedure we used to create
start-goal interpolations with VAEs is not readily available. We
attempted to ``invert'' the generator network using the procedure
suggested in~\cite{creswell_inverting_2016} but our success was
limited since training patters are largely not reproducible by the
generator.
%
%
Although unsuitable for start-goal interpolations, we found that GANS
are very effective to create an autonomous drummer by exploring the
noise space in a smooth way. Exploration can be designed in many ways
and here we propose a very simple approach based on the following
complex periodic function
\begin{equation}
  \label{eq:swirl}
  f(t,\omega_1,\omega_2,\omega_3,\omega_4) \doteq
  e^{\omega_1 jt} -
  \frac{e^{\omega_2 jt}}{2} +
  \frac{je^{\omega_3 jt}}{3}  + \frac{e^{\omega_4 jt}}{4}
\end{equation}
for $t\in[0,2\pi]$ and constants $\omega_1=2$, $\omega_2=19$,
$\omega_3=-20$, $\omega_4=20$. Using a GAN with $d=2$, the real and
the imaginary part of $f$ are used to form the two components of
vector $z$. The resulting ``swirl'' in noise space is illustrated in
Figure~\ref{fig:swirl}.

\section*{Evaluation experiments}

Although patterns generated by VAEs and GANs are novel, we still need
to establish that they do add something new to the current practice of
EDM and that they are of interest to its practitioners. To this end,
we designed three experiments where we asked professional musicians
to assess the quality of the generated patterns.  The
\textit{identification experiment} aims to verify if practitioners are
able to tell start-goal interpolations apart from start-goal
crossfades; the \textit{task experiment} aims to assess how much
musicians appreciated and were able to make use of the drum
interpolation as a compositional tool; the \textit{robot experiment}
aims to rate the aesthetic quality of the autonomous drumming produced by the GAN
when generating patterns by swirling in the representation space.
The goal was to answer the following questions:

\noindent
\textbf{Q1}: Are musicians able to tell interpolations and crossfades
between genres apart during listening sessions?

\noindent
\textbf{Q2}: How do practitioners rate the novelty, adequacy, and
style of the ``instrument'' for creating interpolations between
genres?

\noindent
\textbf{Q3}: Are the drum tracks generated by moving or interpolating
smoothly in the representation space of VAEs and GANs useful as a
material for musicians in composition and performance?

\subsection*{Identification experiment}
The goal of the experiment was to answer Q1.
Subjects were asked to listen to pairs of transitions, a crossfade
and an interpolation. Both \textit{straight} and \textit{mixed} pairs
were formed, in which starting and goal patterns were identical or
different, respectively. Three drum patterns for each of the three
genres were chosen from the dataset. Nine different transitions using
these patterns were specified in a design that includes a transition
for each possible pair of genres in both directions, as well a
transition within each of the three genres. Interpolations and
crossfades had a length of 6 measures (24 bars, 44.7s playing
time). For interpolations, the endpoints were the VAE's
reconstructions of the start and goal pattern.  Crossfades were
produced using a standard function (equal power) of Logic Pro X.

The difference between an interpolation and a crossfade was explained
to the subjects in the visual domain using an animated version of
Figure~\ref{fig:interpolating-vs-xfading}. Every subject was asked to
tell apart 6 pairs, preceded by one practice pair to get acquainted
with the procedure, and received no feedback on the correctness of
their answers.

\subsection*{Task experiment}

The goal of the experiment was to answer Q2 and Q3.  We used the
creative product analysis model (CPAM)~\cite{besemer1999confirming},
that focuses on the following three \textit{factors}: Novelty,
Resolution, and Style. Each factor is characterized by a number of
facets that further describe the product. For each facet, there is a
7-point scale built on a semantic differential: subjects are asked to
indicate their position on the scale between two bipolar words (also
referred to as anchors). Novelty involves two facets: Originality and
Surprise. Resolution considers how well the product does what it is
supposed to do and has four facets: Logicality, Usefulness, Value, and
Understandability. Style considers how well the product presents
itself to the customer and has three facets: Organicness, Well-
craftedness, and Elegance. In this experiment, subjects were allowed
to choose start and goal patterns from those available in the dataset
in order to create their own interpolations using our Ableton Live
interface. In this experiment, subjects were allowed to choose start
and goal patterns from those available in the dataset in order to
create their own interpolations using our Ableton Live interface.

\subsection*{Robot experiment}
The goal of the experiment was to answer Q3.  We used in this case the
Godspeed questionnaire~\cite{bartneck2009godspeed} a well-known set of
instruments designed to measure the perceived quality of robots, based
on subjects' observations of a robot's behavior in a social setting.
They consist of 5-point scales based on semantic differentials.  In
our case, observation is limited to hearing the artificial agent drum
and thus we chose to measure only two factors: Animacy (three facets:
Lively, Organic, Lifelike) and Perceived Intelligence (three facets:
Competent, Knowledgeable, Intelligent).

A long interpolation of 512 bars (124 measures) was generated using
the trained GAN, by ``sweeping'' the code space with a complex
function.
Six segments of 60 bars each were selected from the MIDI file, 9
measures preceded and followed by half a measure (2 bars) for leading
in and out. These MIDI files were rendered into sound using an
acoustic drum soundset in Logic Pro X (Drum Designer/Smash kit), where
the parts of the rimshot and cowbell were transposed to be played by
toms. Acoustic rather than electronic drum sounds were used to
facilitate the comparison with human drumming.
%
Subjects were instructed that they were going to listen to an
improvisation by an algorithmic drummer,
presented with one of the 6 audio files (distributed evenly over the
subject population), and asked to express a judgment on animacy and
perceived intelligence.

\subsection*{Experimental procedure}
The experiments were conducted with subjects active in the wider field
of electronic music (DJs, producers, instrumentalists, composers,
sound engineers), that were familiar with the relevant genres of
EDM\@. Their experience in electronic music ranged from 2--30 years
(median 7 years, average 8.75). They were recruited by the authors
from educational institutes and the local music scenes in Krakow (PL),
Cuneo and the wider Firenze area (IT), and Eindhoven (NL). Experiments
took place in a class room or music studio setting, where subjects
listened through quality headphones or studio monitors. All audio
materials in the experiment were prepared as standard stereo files
(44.1 kHz, 16 bits).

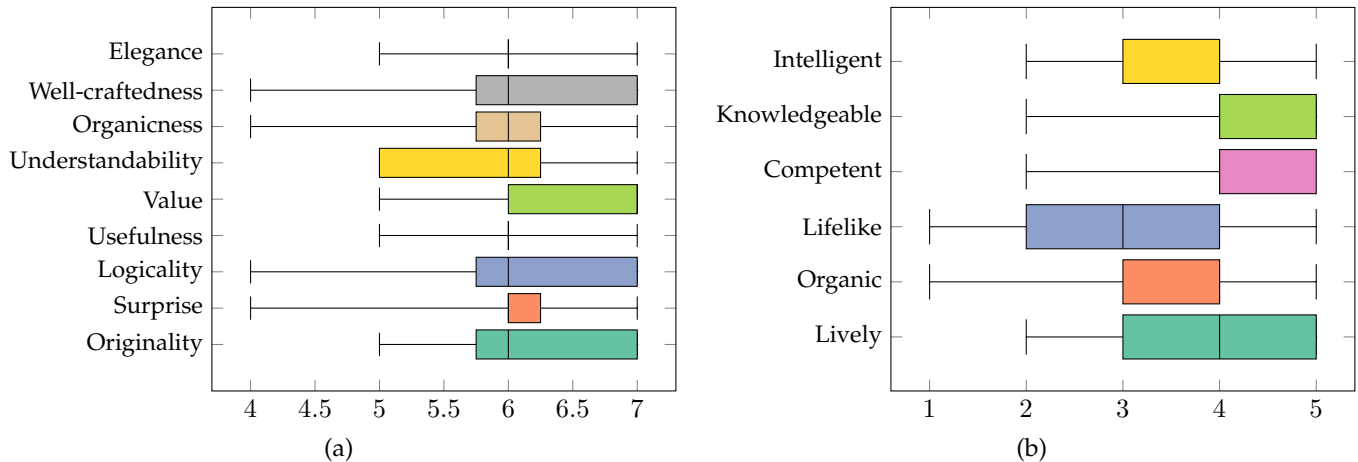
\begin{figure*}
  \begin{tabular}{cc}
    \begin{tikzpicture}
   
      \begin{axis}
        [
        ytick={1,2,3,4,5,6,7,8,9},
        yticklabels={{\small Originality}, {\small Surprise}, {\small Logicality}, {\small Usefulness},{\small Value}, {\small Understandability}, {\small Organicness}, {\small Well-craftedness}, {\small Elegance}},   
        ]
        \addplot+[
        boxplot prepared={
          median=6,
          upper quartile=7,
          lower quartile=5.75,
          upper whisker=7,
          lower whisker=5
        }, fill,draw=black,
        ] coordinates {};
        \addplot+[
        boxplot prepared={
          median=6,
          upper quartile=6.25,
          lower quartile=6,
          upper whisker=7,
          lower whisker=4
        }, fill,draw=black,
        ] coordinates {};
        \addplot+[
        boxplot prepared={
          median=6,
          upper quartile=7,
          lower quartile=5.75,
          upper whisker=7,
          lower whisker=4
        }, fill,draw=black,
        ] coordinates {};
        \addplot+[
        boxplot prepared={
          median=6,
          upper quartile=6,
          lower quartile=6,
          upper whisker=7,
          lower whisker=5
        }, fill,draw=black,
        ] coordinates {};
        \addplot+[
        boxplot prepared={
          median=7,
          upper quartile=6,
          lower quartile=7,
          upper whisker=7,
          lower whisker=5
        }, fill,draw=black,
        ] coordinates {};
        \addplot+[
        boxplot prepared={
          median=6,
          upper quartile=6.25,
          lower quartile=5,
          upper whisker=7,
          lower whisker=5
        }, fill,draw=black,
        ] coordinates {}; 
        \addplot+[
        boxplot prepared={
          median=6,
          upper quartile=6.25,
          lower quartile=5.75,
          upper whisker=7,
          lower whisker=4
        }, fill,draw=black,
        ] coordinates {}; 
        \addplot+[
        boxplot prepared={
          median=6,
          upper quartile=7,
          lower quartile=5.75,
          upper whisker=7,
          lower whisker=4
        }, fill,draw=black,
        ] coordinates {}; 
        \addplot+[
        boxplot prepared={
          median=6,
          upper quartile=6,
          lower quartile=6,
          upper whisker=7,
          lower whisker=5
        }, fill,draw=black,
        ] coordinates {};
      \end{axis}
    \end{tikzpicture}
    &
      \begin{tikzpicture}
        \begin{axis}
          [
          ytick={1,2,3,4,5,6},
          yticklabels={{\small Lively}, {\small Organic}, {\small Lifelike}, {\small Competent},{\small Knowledgeable},{\small Intelligent}},   
          ]
          \addplot+[
          boxplot prepared={
            median=4,
            upper quartile=5,
            lower quartile=3,
            upper whisker=5,
            lower whisker=2
          }, fill,draw=black,
          ] coordinates {};
          \addplot+[
          boxplot prepared={
            median=4,
            upper quartile=4,
            lower quartile=3,
            upper whisker=5,
            lower whisker=1
          }, fill,draw=black,
          ] coordinates {};
          \addplot+[
          boxplot prepared={
            median=3,
            upper quartile=4,
            lower quartile=2,
            upper whisker=5,
            lower whisker=1
          }, fill,draw=black,
          ] coordinates {};
          \addplot+[
          boxplot prepared={
            median=4,
            upper quartile=5,
            lower quartile=4,
            upper whisker=4,
            lower whisker=2
          }, fill,draw=black,
          ] coordinates {};
          \addplot+[
          boxplot prepared={
            median=4,
            upper quartile=5,
            lower quartile=4,
            upper whisker=5,
            lower whisker=2
          }, fill,draw=black,
          ] coordinates {};
          \addplot+[
          boxplot prepared={
            median=4,
            upper quartile=4,
            lower quartile=3,
            upper whisker=5,
            lower whisker=2
          }, fill,draw=black,
          ] coordinates {}; 
        \end{axis}
      \end{tikzpicture}\\
    (a) & (b)
  \end{tabular}
  \caption{(a): Task experiment box plots (n=16, 7-point scale); (b):
    Robot experiment box plots (n= 38, 5-point scale).}
  \label{fig:task-robot}
\end{figure*}

\section*{Results} 
We now present and discuss the experimental results. 

\subsection*{Identification experiment}
This experiment was conducted with 19 subjects using 18 distinct
stimulus pairs. 13 identification errors were made in 114 pairs.  For
each pair correctly identified by a subject 1 point was awarded (0 for
a miss). Subjects achieved an average score of $2.68\pm 0.8$ and
$2.63\pm 0.58$ (out of 3) for straight and mixed interpolations,
respectively. In total they achieved a score of $5.32\pm 1.03$ (out of
6). A Chi-squared test confirms that participants scored better than
chance $\chi^2 (19) = 25.92 $ (critical value $5.99$).  Clearly,
subjects are able to tell interpolations and crossfades apart in a
musical context.

\subsection*{Task experiment}
Fifteen subjects with knowledge of the means of EDM production were
invited to construct an interpolation with the Ableton Live interface
as described above (six of them had previously participated in the
Identification experiment). We asked them to rate their experience
(process and result) on the CPAM scales. Figure~\ref{fig:task-robot}(a)
summarizes the results in a set of box plots, one for each of the
facets. Median scores for all facets are 6 (for \textit{Value} even
7). The average scores for the facets of the factor Resolution
(\textit{Logicality} 6; \textit{Usefulness} 6.13; \textit{Value} 6.5;
\textit{Understandability} 5.8) are generally slightly higher than
those for the factors Novelty (\textit{Originality} 6.13;
\textit{Surprise 5.94}) and Style (\textit{Organic} 5.82;
\textit{Well-craftedness} 6.06; \textit{Elegant} 5.88). Although we
did not use the CPAM to compare different solutions for generating
transitions between drum tracks, subjects judged the process for
creating interpolations and its results against their background
knowledge of existing techniques such as crossfades. The relatively
high scores on all facets indicate that developing the current
prototype into an interpolation instrument will be of value to
practitioners in the field.

\subsection*{Robot experiment}

We asked 38 subjects to listen to a drum track produced by the trained
GAN and to rate the robotic drummer on the scales for Animacy and
Perceived Intelligence.  Figure~\ref{fig:task-robot}(b) summarizes the result
in a set of box plots for the aspects. The median score on all aspects
is 4, with the exception of \textit{Lifelike} where it is 3. Average
scores are higher for the aspects of Perceived Intelligence
(\textit{Competent} 4.24; \textit{Knowledgeable} 3.95;
\textit{Intelligence} 3.84) than for those of Animacy (\textit{Lively}
3.89; \textit{Organic} 3.45; \textit{Lifelike} 3.13). Comments written
by the subjects indicate that they judged Perceived Intelligence
mainly with respect to the construction and evolution of the patterns,
whereas for Animacy the execution of the patterns was more prominent:
absence of small variations in timing and timbre of the drum hits
pushed their judgments towards the anchors Stagnant, Mechanical, and
Artificial. This could be addressed with standard techniques to
``humanize'' sequenced drum patterns by slightly randomizing the note
onsets and velocities, and rotating between multiple samples for each
of the instruments, but for this experiment we used the patterns
output by the GAN without such alterations.  Even though this
measurement just sets a first benchmark for further development, the
high scores for \textit{Competent} and \textit{Knowledgeable} are
encouraging as they suggest that the deep learning process has
captured the genres in the dataset to a large extent.

\section*{Conclusion}
\label{sec:conclusion}


Our tool has already potential applications.  First, it can be used to
improve the process of producing (and delivering) libraries of drum
patterns as the trained network can generate a large number of
patterns in the style represented by the training data.  Second, it
can support the workflows of dance musicians in new ways. Generated
interpolation tracks can be recorded inside the tool to create
fragments to be used in post-production or during live performance as
a foundation on which a DJ or instrumentalist can layer further
musical elements. In addition, VAEs or GANs can be trained on
materials created by individual users, providing users with a highly
customized software instrument that ``knows'' their personal style and
is able to generate new drum tracks in this style for post-production
or in performance.

There are several directions that can be followed to further enrich
the drumming space, including the generation of tempo for tracks that
require tempo that varies over time, and the generation of additional
information for selecting drum sounds in a wide soundset. A more
ambitious direction is to extend our approach for generating whole
sets of instruments (bass lines, leads, pads, etc.) in EDM, which
involves not only note onsets but also pitch and duration.

\bibliography{interpolation,Generative,Evaluation}
\bibliographystyle{IEEEtran}

\end{document}